# STM study of multiband superconductivity in $NbSe_2$ using a superconducting tip

J.G. Rodrigo[*] and S. Vieira

*Laboratorio de Bajas Temperaturas, Departamento de Física de la Materia Condensada,*

*Instituto de Ciencia de Materiales Nicolás Cabrera, Facultad de Ciencias*

*Universidad Autónoma de Madrid, 28049 Madrid, Spain*

**Abstract**

We present a method to produce superconducting tips to be used in Scanning Tunneling Microscopy/Spectroscopy experiments. We use these tips to investigate the evolution of the electronic density of states of $NbSe_2$ from 0.3K up to its critical temperature (7.2K). The use of a superconducting tip (Pb) as ounterelectrode provides an enhancement of the different features related to the DOS of $NbSe_2$ in the tunneling conductance curves, along all the studied thermal range. The analysis of the experimental results gives evidence of the presence of multiband superconductivity in $NbSe_2$.

## 1. Introduction

Today it is widely recognized that scanning tunnelling spectroscopy (STS) is a powerful tool to obtain the local density of states (LDOS) of conducting materials, even at atomic scale. The obtained information is a convolution of the DOS of both electrodes, tip and sample. Therefore, tip preparation and characterization is a crucial point for this technique. Presently there is an increasing interest for the use of superconducting tips, due to the new open possibilities to investigate several important problems previously out of reach [1].

Almost simultaneously, there is a renewed interest in the topic of multiband superconductivity (MBSC), a theory presented by Suhl et al.[2] just two years after the publication of BCS theory. This interest was boosted by the discovery of the superconducting properties of $MgB_2$[3].

The MBSC theory of Suhl et al. predicts that for the case in which two different bands develop superconducting gaps with different values, and there is not interband scattering, two critical temperatures exist. A small amount of interband scattering makes the smaller gap to close at the same $T_C$ as the larger one does. However, there is a temperature interval in which the value of the small gap is reduced close to zero. Recent experiments using ARPES[4] and thermal conductivity[5] techniques strongly suggest that this one could be the case for $NbSe_2$. In order to throw light on this open question tunnel characteristics obtained as a function of temperature could give valuable information, if the region in which the small gap is supposed to decrease is covered with good experimental resolution.

For that reason it is needed to optimise the spectroscopic resolution, mainly for temperatures close to the critical one ($T_C$ = 7.2K), where thermal broadening becomes comparable or larger than the gap values, and the N-S spectroscopic curves become almost featureless. As remarked by Tinkham[6], superconductor-superconductor (S-S) tunnelling provides a very convenient way to determine this quantity because of the sharp features appearing, at a finite temperature, at $|\Delta_1-\Delta_2|$ and $\Delta_1+\Delta_2$ in the current vs. voltage (I-V) characteristics. This situation requires, when using a STM as the spectroscopic tool, to have well characterized superconducting tips. Atomic sharpness is a requisite for the acquisition of topographic images, with atomic resolution, which give valuable information about the quality of the studied region of the sample. To fulfil these requirements it is desirable to have a method for tip preparation and cleaning in-situ, as well as the possibility to do it at will during the experiment at low temperatures.

---
[*] Corresponding author. Tel.: +34 91 497 3800; fax: +34 91 497 3961; e-mail: jose.rodrigo@uam.es

The method of preparation of the tip that we present in this article produces atomically sharp tips with stable and reproducible superconducting properties, and allows a detailed and precised determination of the DOS of the studied samples. In our opinion, this capability may be absent in other methods of preparation of superconducting tips[1].

**2. Experimental**

In this work we present scanning tunnelling microscopy and spectroscopy (STM/S) measurements on $NbSe_2$, in the temperature interval 0.3K<T<7.2K. We have used a home built STM, whose sample holder can move in a controlled way milimetric distances in the x and y directions. The sample was a composite of three different materials, schematically represented in the inset of fig.1. A piece of pure lead (99.999%) was softly flattened by pressing it with a flat ruby, and a previously prepared small piece of gold, around 4 $mm^2$, was embedded into the lead. A small nest was prepared in the lead substrate, and a single crystal of $NbSe_2$ was placed inside. A new and gentle pressing with the ruby crystal forces lead to flow enough as to fill the regions around the crystal and level out the surfaces. Before cooling in an inert atmosphere, the STM tip, made of high purity lead, was cut and placed above the lead region. After cooling down to 4.2 K, the $NbSe_2$ sample was cleaved, leaving a fresh clean surface for experiments. We proceed then with the fabrication of a sharp tip of lead. In several previous articles of our group[7] it has been shown that it is possible to fabricate sharp cone-like nanostructures with a STM at low temperatures. If the tip is deeply indented in a sample made of the same material, in this case lead, a strong contact is formed between them. Then, by means of a controlled pulling and pushing process it is possible to elongate the contact region. The rupture of this contact produces clean atomic sized tips which are stable due to the reduced mobility of lead atoms at the temperatures at which they have been fabricated. Measurements of conductance vs. deformation curves along this process lead to the observation of quite reproducible and characteristic patterns. From these curves, and using simple models[7], it has been possible to obtain a good image of the involved geometries.

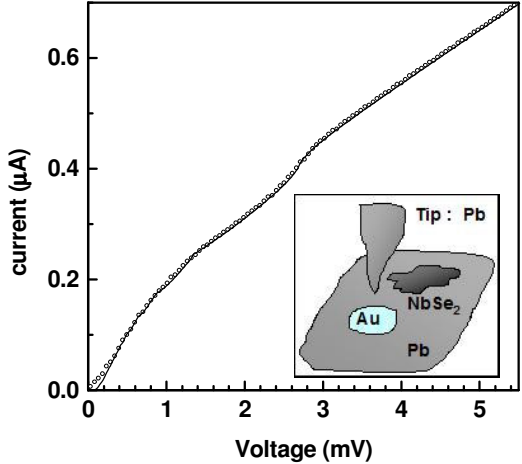

Fig. 1. I-V curve corresponding to a single atom contact (circles), and the fitting using three quantum channels with transparencies 0.87, 0.26 and 0.05 (solid line). Inset: configuration of the composite sample.

Once the tip is created we perform its characterization. This is a two-step process, regarding the atomic sharpness of the tip, and the quality of the spectroscopic I-V characteristics in tunnelling regime, which reflect the superconducting DOS. After the characterization of the tip we will proceed with the study of NbSe2.

**3. Results and discussion**

The first step of the characterization of the tip was done just before breaking the contact, which must be of atomic size. In fig.1 we show a typical experimental curve corresponding to this case. In the quantum transport regime, this kind of non linear curves can be fitted to a sum of contributions from the different quantum channels. The number of channels per atom is a characteristic of the chemical element. It has been shown that for lead this number is three[8]. Non linearity arises from the contribution of Andreev reflection processes to the current. The curve has been fitted following the procedure described in ref.[8]. As it is expected only three channels are needed to obtain the best fit shown in the figure. It is interesting to remark that single atom contacts can be achieved as a routine procedure.

The second step of the characterization is done in tunnelling regime. Conductance vs. applied voltage curves like the one in fig.2(a) are obtained by numerical derivative of the I-V curves. After these measurements, the sample is displaced so the tip is over the gold sample, where S-N conductance curves are taken (fig.2(b)). These steps can be repeated if there is any suspicion about tip sharpness or cleanliness.

In order to perform a detailed analysis of the DOS of the sample, $NbSe_2$, we need to determine the DOS of the Pb tip. Low temperature S-S curves (fig.2 (a)) present a zero value of the conductance inside the gap region. This means that there is not a relevant pair breaking source. In order to account for the finite width and height of the coherence peaks in the conductance curves, we consider the effect of a gap distribution, due to different bands contributing to the Fermi surface in this metal. Multiple gap effects on lead were reported in early experiments[9], and are supported with recent results[10]. As it can be observed in the inset of fig.2, a experimental curve is fitted using for Pb a gaussian distribution of BCS DOS with different $\Delta$, centered at $\Delta_0$=1.375 meV, and with a standard deviation, $\sigma$ = 55 µV. We have also plotted calculated curves using a BCS DOS modified with a lifetime broadening parameter $\Gamma$ [11].

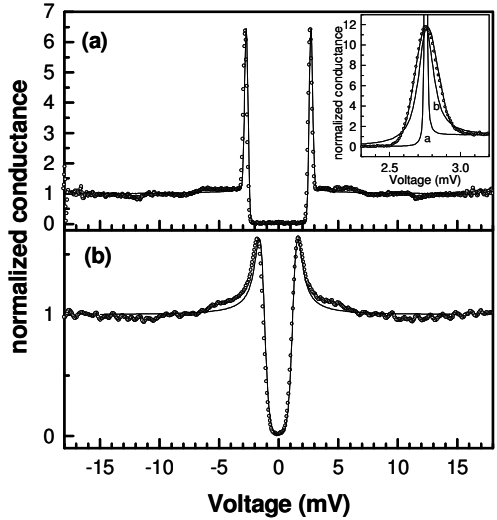

Fig. 2. Pb-Pb (a) and Pb-Au (b) conductance curves in tunneling regime at 2K (dots: experiment; line: fit). Inset: detail of the gap edge fitted using a gaussian distribution, and BCS curves obtained using a single value of $\Delta$ (1.375 meV) and a lifetime parameter, $\Gamma$. Curve (a), $\Gamma$=1µV; curve (b), $\Gamma$=30µV.

It is clear that this model cannot account for the experimental observation. Obviously, the effect of the rms stability in voltage is included in the total width of the gaussian distribution. After analysing thousands of curves, we found that $\sigma$ can vary between 50 and 110 µV, probably due to slight differences in the geometry of the tip apex. Similar gaussian fittings were done for the curves in fig.2.

Once the lead tip is fully characterized we displace the sample holder in such a way that the tip is located over the NbSe$_2$ crystal. In fig.3 (inset) we show a topographic image of its surface. Atomic resolution is easily obtained, and the charge density wave pattern is resolved in many places. Fig.3 shows a typical Pb-NbSe$_2$ conductance curve obtained in tunnelling regime at 4.2 K. The gap related structures at $|\Delta_1-\Delta_2|$ and $\Delta_1+\Delta_2$, and the phonon structures due to the Pb tip are clearly observed.

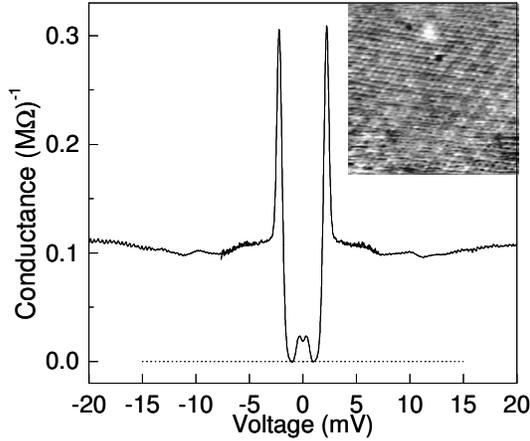

Fig. 3. Pb-NbSe$_2$ conductance curve in tunneling regime at 4.2 K. Inset: 10×10 nm$^2$ STM topographic image of a NbSe$_2$ surface using a Pb tip.

As we will show below, the good knowledge of the tip DOS that was previously achieved, allows to obtain the DOS of the studied material. Next, we discuss our results on the superconducting density of states of NbSe$_2$.

In order to analyze the results, the experimental conductance curves are fitted following the procedure described in ref.[12]. The DOS of NbSe$_2$ is modelled with a phenomenological distribution of gap values,

$$DOS_{NbSe_2} = \frac{1}{\alpha_i} \sum_i \alpha_i \ DOS_{BCS}(\Delta_i)$$

In principle, gap values from 0 to 2 meV are considered in the distribution. Temperature is also included in the model. It is important to indicate that the observation of the main features corresponding to the addition and difference of gap values defines uniquely its central values at each temperature. In the case of lead, these values are in complete agreement with the data that can be found in the literature.

In fig.4 we present experimental curves measured at the same location at 1.9K and 6K, and the fitted conductance curves. In the inset we show the corresponding gap distributions for NbSe$_2$ obtained in the fittings. In some cases it is neccesary to introduce a contribution of non-gapped DOS (i.e., normal state) to the total NbSe$_2$ DOS in order to fit the experimental curves. This contribution becomes important at temperatures above 5K. Slightly different gap distributions can be obtained depending on the location at the sample, or the different geometries of the tip. This kind of fittings have been done in the range of temperatures from 0.4K to 7.2K. From the analysis of the evolution with temperature of the NbSe$_2$ gap distributions presented in fig.5, a picture emerges that supports the MBSC scenario for this compound. At low temperature (below 4 K) the gap distributions extent in a range typically between 0.4 and 1.4 meV, with no indication of gap values close to zero. At higher temperatures the distribution moves to smaller gap values, and the non-gapped contribution increases. Above 5K, the gap distributions clearly present values at zero gap. These observations are in agreement with ARPES experimets at 5.3K [4], indicating that not all the regions of the FS that contribute to the tunnelling curve are fully gapped. This situation is dramatically present at higher temperatures (above 6K) when the non-gapped contribution ammounts more than 40% of the DOS of NbSe$_2$ probed by the Pb tip. In fig.5 we have also sketched the evolution of the gap values of the two main bands of the FS detected in our experiment. The the upper band follows a BCS-like evolution, with a maximum of $2\Delta/k_BT_C \cong 3.9$. On the contrary, in the low limit of the lower band the evolution is clearly non-BCS, following closely the behaviour indicated by Suhl et al.[2] for the case of MBSC with small interband scattering.

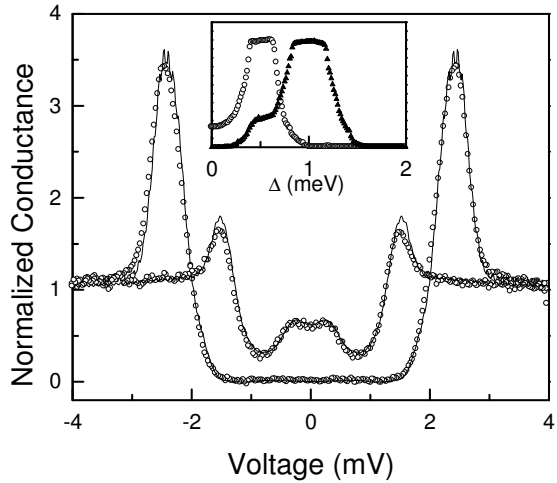

Fig. 4. Pb-NbSe$_2$ conductance curves at 1.9K and 6K (experiment :dots; fitting: line). Inset: gap distributions obtained in the fitting procedure (solid, 1.9K; open, 6K).

The results obtained at low temperatures are compatible with early studies on NbSe$_2$ [13] that showed clearly that a distribution of gap values between 0.7 and 1.4 meV was needed to fit the tunneling experiments.

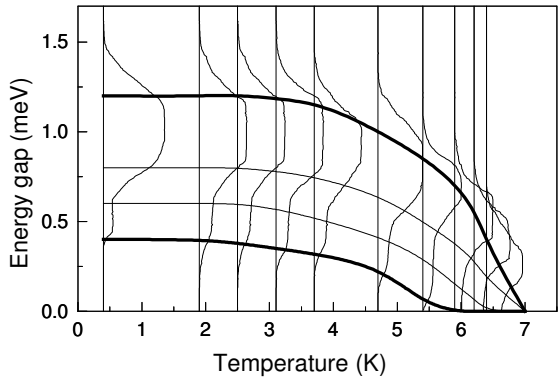

Fig. 5. Evolution with temperature of NbSe$_2$ gap distributions. Lines indicating the evolution of the main features are sketched.

## 4. Conclusions

As a summary, we have presented a new method to create atomically sharp superconducting tips, in-situ, at low temperatures. We have also shown a versatile method for the sample mounting that allows a fast in-situ checking of the quality and characteristics of the superconducting tips. The precise characterization of the apex geometry and DOS of these tips, as well as the absence of uncontrolled effects in them (proximity layer, oxide, etc.), make of them an excellent tool for an exact determination of the DOS of the studied materials. We believe that our tunnelling results strongly support an interpretation in qualitative agreement with the MBSC model by Suhl, Matthias and Walker [2]. More experiments, with a precise determination of the directions in k-space that contribute to the experimental curves, are needed to confirm this picture, but it seems interesting to us that this compound could be a candidate to an old predicted consequence of the MBSC model.

**Acknowledgements**

We wish to thank G. Rubio-Bollinger, H. Suderow and F. Guinea for helpful discussions. Support from the ESF programme VORTEX, DGI-Spain (MAT2001-1281-c02-01) and the Comunidad Autónoma de Madrid, Spain (project: 07N/0039/2002) is acknowledged.